\documentclass[12pt,a4paper]{article}

\usepackage{amssymb,amsfonts}
\parskip=\baselineskip

\begin{document}

\begin{center}
{\large\bf\uppercase{On the universal X-ray 
luminosity function of binary X-ray sources in
galaxies\\}}
\vskip\baselineskip

{\large\bf
K.\,A.\,Postnov
}\\

{\it Sternberg Astronomical Institute, Moscow University, 
Russia}
\end{center}

\vskip\baselineskip

\vskip\baselineskip

{\it Abstract}. 
The empirically determined universal power-law shape of 
X-ray luminosity function of high mass X-ray binaries in 
galaxies is explained by fundamental mass-luminosity and 
mass-radius relations for massive stars.  
\vskip\baselineskip

The unique possibilities offerred by modern X-ray space observatories 
{\it Chandra} and {\it XMM-Newton} make it possible to observe 
point-like X-ray sources in nearby galaxies
which allows studies of formation and evolution 
of the population of close binary systems as a whole. 
Recently, Grimm, Gilfanov and Sunyaev (2002, 2003) 
reported that the differential luminosity function 
of point-like X-ray sources in nearby galaxies
has a universal power-law shape 
$$
\frac{dN}{dL_x}\propto \hbox{SFR}\times L_x^{-\alpha}
$$  
in the wide range of X-ray luminosities 
$\sim 10^{34}-10^{40}$ erg/s, where SFR means the star
formation rate in a specific galaxy, 
$\alpha=1.61\pm 0.12$. 

In this Letter we show that the empirically found power-law form
of the X-ray luminosity function naturally appears during accretion 
of matter in close binary systems. In high-mass X-ray binaries,
the compact star accretes matter from stellar wind from an early-type
optical companion. Observations suggest (De Jager, 1980) that 
stellar wind mass loss rate from massive ($>$ several  $M_\odot$) 
main sequence stars can be parametrized as 
\begin{equation}
\dot M_{w}\propto \frac{L_o}{v_\infty}
\label{LF}
\end{equation}
where $L$ is the stellar luminosity,   $v_\infty$ is the stellar
wind velocity at the infinity. With a good accuracy the latter is 
proportional to the escape velocity near stellar photosphere 
$v_\infty\approx 3v_p$. Formula (\ref{LF}) simply refelcts 
the radiation pressure driven wind. 

It is well known that for stationary stars on the main sequence
power-law mass-luminosity (M-L) and mass-radius (M-R) relations hold 
(see e.g. Zel'dovich, Blinnikov, Shakura 1981). The M-L relation 
follows from the diffusive character of the 
heat transfer from stellar interiors 
\begin{equation}
L\propto 1/\kappa\rho \nabla T^4
\label{diff}
\end{equation}
($\kappa$ is the opacity, $\rho$ is the density, 
$T$ is the temperature) with account of the 
virial relation $T\propto M/R$ for stationary stars.
For sufficiently massive stars Tompson scattering dominates
in the opacity and $L\propto M^3$
(Eddington, 1926). The empirically determined power-law index
for early-type stars is close to the theoretically expected value 3,
with slightly decreasing down to 2.7 for very massive stars
(De Jager, 1980).  

The M-R relation can be obtained by using the energy release per gramm
in the stellar core due to thermonuclear reactions
\begin{equation}
\epsilon\sim \rho T^{Ze}
\end{equation}
where $Ze=\partial\log <\sigma v_0>/\partial \log T$ is the 
Zel'dovich number showing the temperature dependence of thermonuclear
reactions (see Blinnikov 2000) and radiation diffusion equation  
(\ref{diff}): $R\propto M^{Ze-1/Ze+3}$. 
For example, for the pp-cycle $Ze\approx 3...10$, 
for the CNO-cycle $Ze\approx 10...27$, depending on the central 
temperature. The empirically found mass-radius relation 
for main sequence stars is $R\propto M^{0.8}$ with some 
decrease in the power-law index for very massive stars.

The accretion rate onto a compact star from stellar wind 
$\dot M_a$ is determined by its orbital velocity 
($v_{orb}$), the masses by the binary system companions  
($M_x$ and $M_o$ for the compact and optical star, respectively) 
and by the stellar wind velocity
(Lipunov 1992). In the case of fast stellar wind 
($v_w>v_{orb}$), the Bondi-Hoyle accretion rate is
$$
\dot M_a\propto \dot M_w \frac{M_x^2}{a^2 v_w^4}
$$
(here $a$ is the binary's semi-major axis). 
Substituting equation (\ref{LF}) for stellar wind mass loss
rate and using M-L and M-R relations yield the X-ray luminosity 
$$
L_x\sim \dot M_a\sim M_o^{2.5}/a^2\,. 
$$
Observations also suggest that the binary systems semi-major
axis distribution has a flat shape $d\log a=const$
(Massevich, Tutukov 1988) and thus is independent of 
the optical star mass, so $M_o\sim L_x^{2/5}, \quad 
dM_o/dL_x\sim L_x^{-3/5}$. 

Then for the luminosity function of X-ray sources accreting from
stellar wind,
$$
\frac{dN}{dL_x}=\frac{dN}{dM_o}\frac{dM_o}{dL_x}\,,
$$
using the Slapetdr mass function for main sequence stars
$dN/dM_o\sim M_o^{-2.35}$, we get
\begin{equation}
\frac{dN}{dL_x}\sim L_x^{-0.94}L_x^{-3/5}\sim L_x^{-1.54}
\label{dNdL}
\end{equation}
which is very close to what is actually observed.
In fact, in the stationary case, 
instead of using the Salpeter initial mass funciton, 
we should utilize the steady-state mass distrubution with account 
of the mass loss rate  
$dM_o/dt\sim M_w$, which would make the mass dependence even 
steeper than the initial mass function has, so the power-law index 
in equation (\ref{dNdL}) changes towards higher values,
as we need. 

In the case of slow stellar wind
$v_w<v_{orb}$, the rate of mass capture by the compact star
is mainly determined by its orbital velocity, which for 
high-mass X-ray binaries with 
$M_o/M_x >>1$ depends on the optical star mass 
in the similar way as the escape velocity does, 
so no qualitative changes are expected.

Note that although the chemical composition is important 
for absolute values of stellar luminositites and radii, 
it does not affect their functional dependence on mass
which determines the shape of the luminosity function.

Remarkably, a close to the obtained dependence can be formally
derived in the case of accretion from optical star through the  
Roche lobe overflow, too. Indeed, in this case the accretion rate
occurs in the thermal time scale of normal star, $\tau_{KH}$:  
$\dot M_o\sim M_o/\tau_{KH}\sim M_oR_oL_o/GM_o^2$.
Then again $L_x\sim \dot M_o\sim M_o^{2.8}$, $M_o\sim L_x^{0.36}$, 
$$
\frac{dN}{dL_x}\sim L_x^{-0.94}L_x^{-0.64}\sim L_x^{-1.58}\,.
$$  
However, in real high-mass X-ray binaries the mass 
transfer rate in the thermal time scale leads to the 
supercritical accretion regime, the accretion luminosity 
becomes on the order of the Eddington one and does not
depend on the optical star mass. In reality, in stationary
case it can be notably less than the Eddington value
due to absorption in the powerful wind outflow from
the accretion disc, as in the case of SS 433. 
Such sources are not numerous and can not significantly affect 
the shape of the X-ray luminosity function.
\vskip\baselineskip

I acknowledge R.S.Sunyaev and M.R.Gilfanov for discussions.
The work is partially supported by the RFBR grant 
00-02-17164.
 
\vskip\baselineskip

\end{document}